Title: Ultra-thin strain-relieving $Si_{1-x}Ge_x$ layers enabling III-V epitaxy on Si

Authors: Trevor R. Smith, Spencer McDermott, Vatsalkumar Patel, Ross Anthony, Manu Hedge,[a] Andrew P. Knights, Ryan B. Lewis*

Department of Engineering Physics, McMaster University, L8S 4L7 Hamilton, Canada

[a]Present address: Air Liquide, 19702, Newark, USA

*Email: rlewis@mcmaster.ca

Statement on Ethics and Integrity Policy: Financial funding of this project is from the Natural Sciences and Engineering Research Council of Canada under projects RGPIN-2020-05721 and ALLRP-588298-23, and the National Research Council's High Throughput Secure Networks Challenge Program. There are no conflicts of interest with the authors of this manuscript.

Abstract: The explosion of artificial intelligence, possible end of Moore's law, dawn of quantum computing and continued exponential growth of data communications traffic have brought new urgency to the need for laser integration on the diversified Si platform. While diode lasers on III-V platforms have long powered internet data communications and other optoelectronic technologies, direct integration with Si remains problematic. A paradigm-shifting solution requires exploring new and unconventional materials and integration approaches. In this work, we show that a sub-10-nm ultra-thin $Si_{1-x}Ge_x$ buffer layer fabricated by an oxidative solid-phase epitaxy process can facilitate extraordinarily efficient strain relaxation. The $Si_{1-x}Ge_x$ layer is formed by ion implanting Ge into Si(111) and selectively oxidizing Si atoms in the resulting ion-damaged layer, precipitating a fully strain-relaxed Ge-rich layer between the Si substrate and surface oxide. The efficient strain relaxation results from the high oxidation temperature, producing a periodic network of dislocations at the substrate interface, coinciding with modulations of the Ge content in the $Si_{1-x}Ge_x$ layer and indicating the presence of defect-mediated diffusion of Si through the layer. The epitaxial growth of high-quality GaAs is demonstrated on this ultra-thin $Si_{1-x}Ge_x$ layer, demonstrating a promising new pathway for integrating III-V lasers directly on the Si platform.

The local, monolithic integration of semiconductor lasers with Si electronic and optical components would revolutionize data communication and computing hardware architectures. Direct bandgap III-V compound semiconductor optoelectronics—the workhorses of the data communication industry—present obvious appeal for integration in the Si platform. While many approaches for integrating lasers on Si have been explored, all face drawbacks[1–3]. Hybrid integration techniques, such as micro-transfer printing, flip-chip bonding and photonic wire bonding, involve the transfer bonding or coupling of prefabricated III-V lasers. While the devices are high quality, the techniques require each component to be individually aligned, limiting scalability[4,5]. Heterogeneous integration involves wafer-scale bonding and epitaxial layer lift-off, but has extreme process requirements and typically results in III-V devices with limited performance[6,7]. All the above approaches are costly and pose challenges for wafer-scale laser integration with Si-based photonic integrated circuits (PICs). The direct growth of III-V semiconductors on Si is conceptually the simplest integration approach, but in reality, has proven the most difficult. For direct III-V growth on Si, challenges include minimizing threading dislocation densities (TDD), overcoming lattice and thermal mismatches between materials and the polar/non-polar III-V/IV interface, which leads to anti-phase domains (APDs).

To manage the above issues, direct-growth typically employs several-micron-thick dislocation-filtering buffer layers[8]. These layers pose several problems: 1) thermal mismatch between the III-V and Si can create defects and cracks as the substate is cooled post-growth[9]. 2) The layers are optically absorbing, which makes it difficult to couple light to the underlying Si structures[10]. The direct heteroepitaxy of III-Vs has been more successful on Ge than on Si, with (In,Ga)P/(In,Ga)As/Ge multijunction photovoltaics being commercially available and GaAs-based lasers demonstrated[11]. The success on Ge results from the near-perfect GaAs–Ge lattice matching—the Ge lattice is 0.08% larger than that of GaAs—as well as similar thermal expansion coefficients [11–15]. However, GaAs growth on Ge—and III-V-on-IV heteroepitaxy in general—must contend with APDs due to the polar/non-polar interface, which can be minimized on both (100) and (111) orientations with appropriate growth optimization. Specifically, on Ge(100) substrates offcut towards (011), APDs self-terminate when Ga-Ga and As-As anti-phase boundaries meet as they propagate along the (111) and (1$\bar{1}\bar{1}$) bond planes. The As/Ge interface in GaAs/Ge(111) preferentially absorbs onto sites with one dangling bond, resulting in a predominantly (111)B orientation[16–18]. While APD-free GaAs-on-Ge can be achieved, persistent high TDD remain detrimental for laser performance, resulting in the formation of non-radiative dark-line defects[19,20]. Careful optimization of growth conditions is required to achieve high quality GaAs on Ge, with key growth parameters including growth temperature and buffer-layer thickness[21], substrate offcut[22], and AsH$_3$ partial pressure[23]. We note that most work has focused on the (100) surface.

One approach for III-V epitaxy on Si is to use a Ge buffer layer. The Ge layer is typically grown several microns thick by vapor-phase epitaxy to relax misfit strain and bury defects[24]. However, a unique solid-phase epitaxy (SPE) process for $Si_{1-x}Ge_x$ on Si has also been reported, which involves forming an amorphous $Si_{1-x}Ge_x$ layer, followed by selective oxidation of Si and SPE condensation of a Ge-rich single crystal layer[25–31]. However, the detailed Si-Ge misfit strain relaxation mechanisms for this unique SPE process, as well as the employment of this technique for III-V epitaxial growth have not been investigated.

In this work, we show that sub-10-nm-thick strain-relieving $Si_{1-x}Ge_x$ layers can be realized by Ge ion implantation and selective oxidation of Si(111) wafers. The resulting Ge-rich layers are fully strain relaxed via a network of misfit dislocations at the Si-$Si_{1-x}Ge_x$ interface, which do not propagate through the $Si_{1-x}Ge_x$ film. The dislocation network coincides with periodic composition variations at the Si-$Si_{1-x}Ge_x$ interface—the result of the defect-medicated diffusion of Si atoms from the Si substrate through the $Si_{1-x}Ge_x$ layer during oxidation. The epitaxial growth of GaAs on these ultra-thin virtual substrates is demonstrated, presenting a promising approach for solving the long-standing challenge of local, monolithic integration of III-V optoelectronics on the Si platform.

The GaAs/$Si_{1-x}Ge_x$/Si(111) heterostructure fabrication process is outlined in Figure 1a. The process first entails implanting Si(111) substrates with a high fluence of 30 keV Ge+ ions— $7.50 \times 10^{15}$ cm$^{-2}$ (sample A) and $2.25 \times 10^{16}$ cm$^{-2}$ (sample B). Previous investigations of Ge-implanted Si at similar conditions have shown the implantation to result in an amorphous $Si_{1-x}Ge_x$ surface layer[25,28]. After implantation, a 900 °C wet oxidation is carried out to preferentially oxidize Si atoms—resulting in a Si-rich surface oxide—and to promote the condensation and recrystallization of a Ge-rich $Si_{1-x}Ge_x$ interface layer between the oxide and the underlying Si substrate. The surface oxide is subsequently removed, and the substrate used for GaAs heteroepitaxy by organo-metallic vapor-phase epitaxy (OMVPE).

Figures 1b–g show cross-sectional high-angle annular dark-field scanning transmission electron microscopy (HAADF-STEM) images and corresponding energy dispersive spectroscopy (EDS) maps of the two GaAs/$Si_{1-x}Ge_x$/Si(111) heterostructures in the vicinity of the $Si_{1-x}Ge_x$ interfacial layer. For both samples, the Ge is confined to a thin (< 10 nm) region between the Si substrate

and the GaAs epilayer. For sample A (Figures 1b–d)—containing the lower Ge dose—the Ge interfacial layer is non-uniform, exhibiting Ge-rich regions. These clusters presumably formed during the high-temperature/high-diffusion oxidative SPE process. The Ge profile for sample A has a mean thickness of 3.21 ± 1.63 nm, with a variance to mean ratio (VMR) of 0.51. In contrast, sample B (Figures 1e–g)—containing the higher Ge dose—exhibits a highly uniform distribution of Ge in the $Si_{1-x}Ge_x$ interlayer. The Ge profile has a mean thickness of 5.77 ± 0.54 nm, with the VMR of 0.09. Low-magnification TEM micrographs of the entire structure, as well as high-resolution x-ray diffraction (HR-XRD) measurements indicate higher crystal quality of the GaAs for sample B (see Supporting Information). In both cases, the GaAs layers are single-orientation and fully strain relaxed. Given the near-perfect lattice matching of Ge to GaAs, a single-crystal $Si_{1-x}Ge_x$ layer with near-unity Ge content and full strain relaxation is the ideal starting surface for GaAs heteroepitaxy. We note that these conditions are not achievable for such thin crystalline layers deposited on Si using conventional growth techniques.

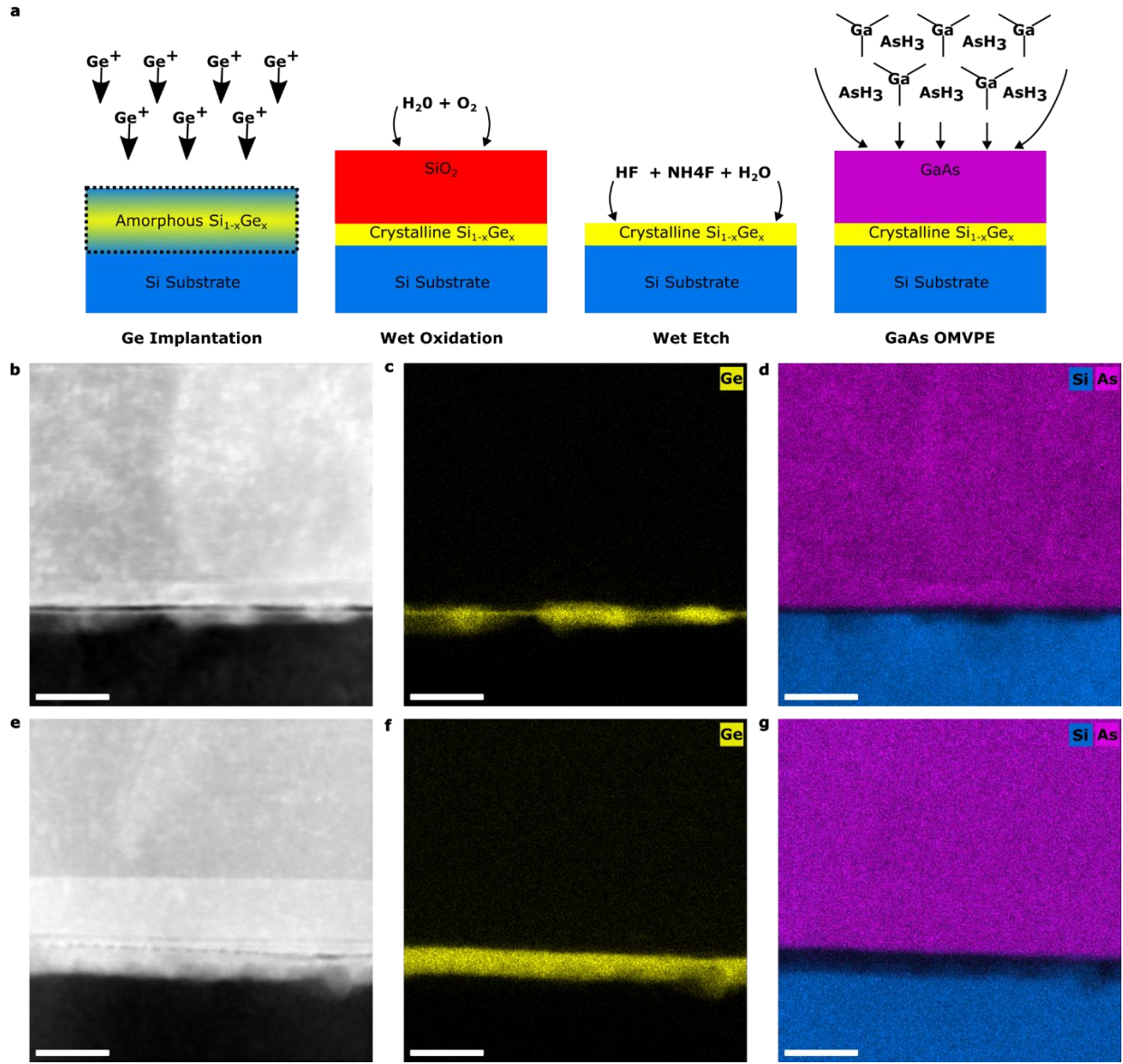

*Figure 1: a) Schematic of the sample fabrication process. First, the implantation of Ge ions into the substrate; the implantation results in an amorphous $Si_{1-x}Ge_x$ surface layer. Second, the wet oxidation of Si and the SPE of Ge-rich $Si_{1-x}Ge_x$ layer. Third, removal of surface oxide with a wet etch. Lastly, the growth of GaAs on $Si_{1-x}Ge_x$ by OMVPE. b) cross-sectional HAADF-STEM image of sample A. c–d) Sample A EDS maps of Ge, and As and Si, respectively. e) Cross-sectional HAADF-STEM image of sample B. f–g) Sample B EDS maps of Ge, and Si and As, respectively. In the EDS maps, Ge, As and Si are shown as yellow, blue and magenta, respectively. Scale bars correspond to 20 nm.*

To explore the strain and interface properties of sample B, atomic-resolution HAADF-STEM experiments were carried out. Figure 2a presents a cross-sectional image along $\langle 11\bar{2} \rangle$ (see Supporting Information for corresponding Fast Fourier Transforms). Figure 2a exhibits an intriguing arch-like periodic contrast variation in the $Si_{1-x}Ge_x$ layer along the $Si_{1-x}Ge_x$/Si interface. As will be confirmed below, this contrast variation corresponds to a periodic modulation of the Ge content in the $Si_{1-x}Ge_x$ layer near the Si interface—darker regions having lower Ge content. To our knowledge, such a structure has not been previously reported. The arch interface structure is of particular interest due to the regular periodicity and geometry. At the upper $Si_{1-x}Ge_x$/GaAs interface, local regions of dark contrast are observed, which coincide with the presence of oxygen observed in EDS (not shown)—possibly a result of incomplete oxide removal before GaAs deposition. To further examine the arch-like structure at the interface, Figure 2b shows a STEM micrograph taken along a ⟨11$\bar{1}$⟩ inclined from the interface by 19.5°. This image exhibits multiple rows of arches along the $Si_{1-x}Ge_x$/Si interface at different heights—depths along the [11$\bar{2}$] direction. Each row of arches is spaced laterally by approximately twice that seen in Figure 2a. As the rows of arches are offset by half a period, a single arch pattern with half the spacing is observed when viewed along the [11$\bar{2}$] direction in the interface plane (c.f. Figure 2a).

Geometric phase analysis (GPA) strain maps of Figure 2a are shown in Figures 2c–d, for g-vectors of $g_1$ = [2$\bar{2}$0] and $g_2$ = [111], corresponding to $\epsilon_{xx}$ and $\epsilon_{xx}$ strain, respectively. Strain dipoles in the $\epsilon_{xx}$ GPA image illustrate the network of dislocations at the $Si_{1-x}Ge_x$/Si interface, which are verified to be present in the STEM image (see Supporting Information). The location of the dislocations aligns with the bottom of the arches. We note the GPA maps have been cropped to remove edge artifacts (see Supporting Information for full GPA maps). Assuming a fully-relaxed Si substrate as a reference, the GPA data in Figures 2c–d indicate a $Si_{1-x}Ge_x$ average lateral and vertical lattice constants of 5.59 Å and 5.67 Å, respectively. We note that a single dislocation is observed in Figure 3c near the GaAs/$Si_{1-x}Ge_x$ interface—otherwise, the GaAs/$Si_{1-x}Ge_x$ interface is coherent.

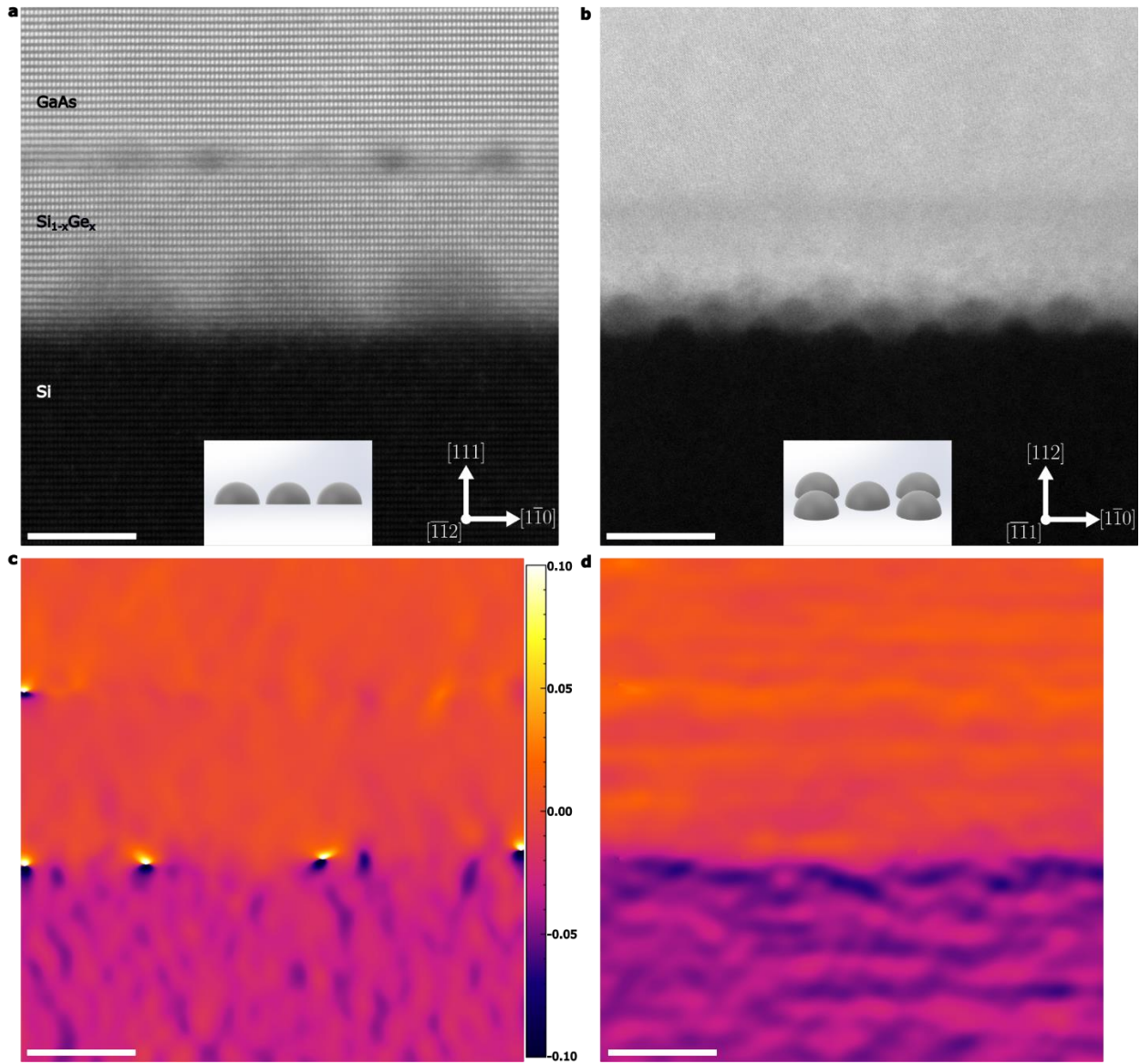

*Figure 2: HAADF-STEM images and GPA maps of the Si/Si$_{1-x}$Ge$_x$/GaAs heterostructure (sample B). a) STEM image along [11$\bar{2}$] with inset illustrating the periodic arches. b) STEM image along [11$\bar{1}$] (inclined from the interface) with inset illustrating periodic arches. c–d) $\epsilon_{xx}$ and $\epsilon_{yy}$ GPA strain maps of micrograph a). The colorized scale bar applies to c–d. Scale bars correspond to 5 nm in a, c, and d, and 20 nm in b.*

Figure 3 explores the local composition profile of the Si$_{1-x}$Ge$_x$ arches of sample B. Figures 3a–d present EDS maps of Ge and Si, and a corresponding HAADF image around the Si$_{1-x}$Ge$_x$ arches. The composition maps reveal that the arches correspond to a periodic composition modulation of the Si$_{1-x}$Ge$_x$ layer. EDS line scans of Ge content are presented in Figure 3e at different heights in the Si$_{1-x}$Ge$_x$ layer, as indicated in Figure 3d. The atomic composition of Ge increases as the scan height increase from the Si$_{1-x}$Ge$_x$/Si interface to the GaAs/Si$_{1-x}$Ge$_x$ interface, reaching a maximum of roughly 80% near the GaAs/Si$_{1-x}$Ge$_x$ interface. The dark area in the arch structure in Figure 3d corresponds to regions of higher concentration for Si, as seen in the oscillations in concentration along the $Si_{1-x}Ge_x$ layer of Figures 3a–c. We note that $Si_{0.2}Ge_{0.8}$ corresponds to a calculated strain-free lattice constant of 5.61 Å, similar to the average lattice constant measured above by GPA. The arch structure has an average period of 7.7 ± 0.5 nm.

Strain-relaxation via dislocations in $Si_{1-x}Ge_x$/Si films normally occurs via 60° ⟨110⟩ misfit dislocations[32]. For (111) heterostructures, there are three active {111} glide planes—{11$\bar{1}$}, {1$\bar{1}$1}, {$\bar{1}$11}—with corresponding dislocation lines running along [1$\bar{1}$0], [10$\bar{1}$] and [01$\bar{1}$], respectively— the intersection of the glide planes with the {111} interface. Given the Si lattice constant of 0.543 nm and relaxation occurring over all 3 glide planes, an arch/dislocation spacing of 7.7 nm would accommodate 3.7% of lattice mismatch. Thus, this dislocation network is sufficient to fully relax the misfit strain of a $Si_{1-x}Ge_x$ layer with nearly 90% Ge. Furthermore, the correspondence of the dislocations with the Ge composition modulation implies that the dislocations formed during the initial oxidative SPE process, indicating this unusual growth mode is extremely efficient at forming dislocations to relax strain, likely a result of the high temperature and large diffusivity of atoms involved.

The correspondence of concentration oscillations in the $Si_{1-x}Ge_x$ film with the network of interface dislocations is indicative of a defect-mediated diffusion process occurring during the $Si_{1-x}Ge_x$ SPE process. A schematic of this defect-mediated diffusion is shown in Figure 3f. During the SPE process, Si diffuses up toward the oxide, replacing Ge at the oxide interface. In turn, Ge diffuses downward as the $Si_{1-x}Ge_x$ layer propagates into the Si substrate. The supply of Si to the oxide is maintained throughout the oxidation by the Si substrate, which diffuses upward through the $Si_{1-x}Ge_x$ layer to the oxide interface. As the oxide thickness increases during annealing, the $Si_{1-x}Ge_x$ layer moves downward. The Si–Ge exchange from diffusion at the oxide–layer interface and layer–substrate interface allows for the $Si_{1-x}Ge_x$ to persist as it diffuses deeper. The results presented in Figures 2 and 3 indicate that in the vicinity of the dislocations, there is a relative tensile strain in the layer and a reduction in the local Ge concentration. During the oxidation of the implanted substrate, Si atoms diffuse up from the substrate through the Ge-rich layer to be oxidized. The lack of oxygen below the $Si_{1-x}Ge_x$ layer indicate that it is Si that diffuses up through the layer, and O does not cross the layer. The coincidence bottom of the $Si_{1-x}Ge_x$ arches with the dislocations indicates that diffusion happens most rapidly in the vicinity of the dislocations, as the Ge interface is deeper into the substrate in these locations. This effect results in the arch shape of the $Si_{1-x}Ge_x$ layer. This novel high temperature process facilitates the relaxation of misfit strain with extremely high efficacy.

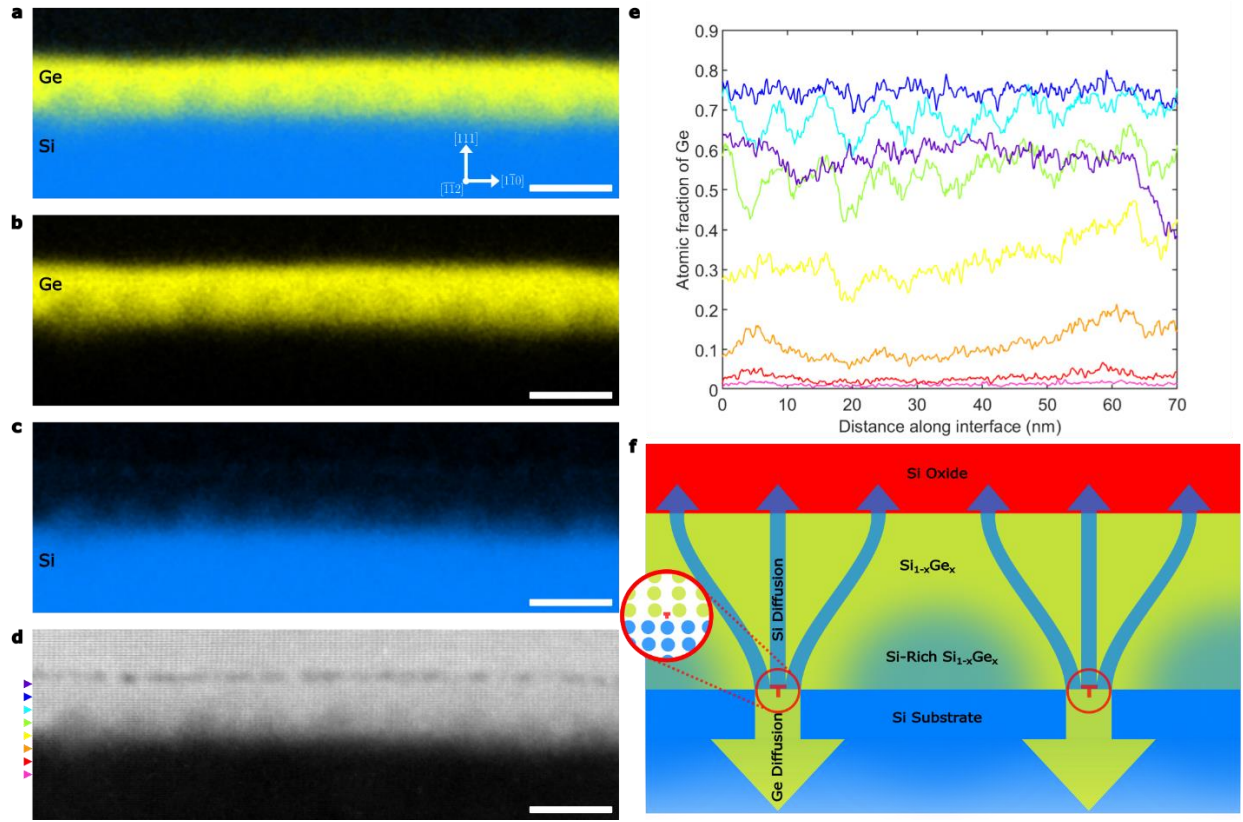

*Figure 3: a–c) EDS maps of Si and Ge, Ge, and Si, respectively, from sample B. d) HAADF micrograph of the EDS scan region. e) Line scans of Ge concentration taken at various locations between the to $Si_{1-x}Ge_x$ interfaces, as indicated by the colored markings in d). f) Schematic of the defect-mediated diffusion process, red circles around the edge dislocation symbol indicate dislocations and arrows represent diffusion through the defects at the $Si_{1-x}Ge_x$–substrates interface: yellow arrows show Ge diffusion downward into the substrate, and blue arrows represent Si diffusion from the defect and into the oxide. Inset shows illustration of dislocation at Si–Ge interface Scale bars in a–d is 10 nm.*

Making the direct growth of III-V lasers on Si a reality will take new scientific innovations. We have demonstrated a new approach for the growth of GaAs and related III-V technologies on the Si platform via a sub-10-nm-thick strain-relaxed $Si_{1-x}Ge_x$ buffer layer, fabricated by an unusual oxidative solid-phase epitaxy process. The unique growth process relaxes the $Si_{1-x}Ge_x$/Si misfit strain with remarkably efficiency, producing a network of dislocations at the $Si_{1-x}Ge_x$/Si interface, along with corresponding periodic composition variations that are the result of a novel defect-enhanced adatom diffusion process. The defect-mediated diffusion in the $Si_{1-x}Ge_x$ layer during oxidation is—to our knowledge—a novel phenomena. These results present an exciting platform for the III-V heteroepitaxy on Si, which could enable the direct growth and integration of viable III-V lasers with Si photonic integrated circuits and microelectronics, enabling the next generation of semiconductor chips for data, computing and quantum applications.

ACKNOWLEDGEMENTS

The authors are grateful to Carmen Andrei and Natalie Hamada for TEM support. We acknowledge support from McMaster University's Centre for Emerging Device Technologies and the Canadian Centre for Electron Microscopy. We are grateful for financial support from the Natural Sciences and Engineering Research Council of Canada under projects RGPIN-2020-05721 and ALLRP-588298-23, and the National Research Council's High Throughput Secure Networks Challenge Program.

METHODS

Si(111) substrates were implanted with $7.50 \times 10^{15}$ Ge$^+$cm$^{-2}$ (sample A) and $2.25 \times 10^{16}$ Ge$^+$cm$^{-2}$ (sample B) at 30 keV. After implantation, the samples were wet-oxidized at 900 °C for 30 minutes to form the Si$_{1-x}$Ge$_x$ layer through the preferential oxidation of Si and recrystallization of the ion-induced amorphized Si$_{1-x}$Ge$_x$. Prior to loading the samples for OMVPE growth, the SiO$_2$ surface oxide was removed with a 10:1 buffered oxide etch. OMVPE growth of GaAs was subsequently carried out directly on the Si$_{1-x}$Ge$_x$ surface layers. TEGa and AsH$_3$ were used as precursors with flow rates of 150 sccm and 1.85 sccm, respectively. The growth was conducted for 10 minutes at a substrate temperature of 630 °C and pressure of 100 Torr in a close-coupled showerhead reactor from Structured Materials Industries.

The GaAs/Si$_{1-x}$Ge$_x$/Si samples were characterized using cross-sectional STEM and HR-XRD to assess film quality. All samples were prepared for TEM using a Helios 5 UC DualBeam by Thermo Scientific. A 300-nm-thick layer of carbon was deposited for surface protection from the ion beam with an electron beam, and 3.3-um-thick layer of W was deposited with an ion beam. The TEM samples then underwent a cleaning process with exposure of the TEM lamella to a 2 kV, 0.19 nA Ga focused-ion beam (FIB) for several minutes on each side. STEM was done on two different instruments; A Talos 200X was used for HAADF-STEM imaging (Figure 1), operating at 200 keV with four in-column EDS detectors using a double-tilt holder. All other STEM images presented were collected with a FEI Titan 80-300 LB microscope in HAADF-STEM, operating at 300 keV with a hexapole aberration corrector, monochromator, and EDS detector using a double-tilt holder.

EDS maps presented show the normalized atomic composition for selected elements in the electron micrograph. TEM and EDS micrographs were acquired and processed in the Thermo Fisher Velox software. For EDS atomic concentration analysis, the concentration of Si and Ge in films were spatially determined by collecting several lateral concentration profiles at various vertical positions throughout the film. Each concentration value was filtered by averaging over 10 pixels vertically (approximately 1.5 nm). Strain analysis was performed for the identification of dislocations at the heteroepitaxy interface by geometric phase analysis (GPA) with perpendicular g-vectors using Strain++. Si$_{1-x}$Ge$_x$ film thickness was extracted from EDS micrographs using the average full width half maximum (FWHM) of the Ge concentration profile, with the variance over mean ratio (VMR) used to quantify uniformity. The VMR metric is bounded between 0 and 1, as a VMR of 0 is perfectly uniform and 1 is completely random.

HR-XRD reciprocal space mapping was performed using a Rigaku SmartLab system with a K-$\alpha_1$ wavelength Cu source for x-ray generation. A 4-bounce Ge monochromator was used for the incident beam, and a 2-bounce analyzer crystal for the diffracted beams. Maps were collected around the (440) peaks of Si and GaAs/Ge.

# Supporting Information

This supporting information provides additional transmission electron microscopy (TEM) images/analysis and high-resolution x-ray diffraction (HRXRD) characterization of the two samples investigated in the main text—samples A and B implanted with $7.50 \times 10^{15}$ cm$^{-2}$ and $2.25 \times 10^{16}$ cm$^{-2}$ Ge, respectively. Additional details about the geometric strain analysis (GPA) characterization of TEM images is also given.

## XRD

Figure S1a-b presents cross-sectional TEM micrographs of samples A and B. Sample A shows greater thickness variation compared to that of sample B with substantially more variance visible at the surface of the GaAs sample, more grain boundaries in the GaAs layer, and increased contrast variations at the GaAs/SiGe interface—implying a higher density of defects and lower overall crystal quality for sample A. This supports the notion that the lack of uniformity of the SiGe layer increases interfacial strain leading to defective crystal growth. Although sample B is smoother than A, both exhibit significant surface roughness. Notably, GaAs(111) homoepitaxy is sensitive to growth conditions and we believed that improved growth condition or the use of surfactants may lead to a more uniform film. We note that GaAs(111) homoepitaxy requires careful optimization of growth conditions to achieve smooth surfaces. It is expected that smoother GaAs surfaces could have been obtained for sample B after optimization of the GaAs(111)A MOVPE growth conditions.

Figure S1c-d shows high-resolution x-ray diffraction reciprocal space maps (RSMs) around the (440) GaAs and Si peaks for samples A and B. In both cases, a peak is observed at the location of 100% relaxed GaAs/Ge, indicating that the GaAs layers are fully strained relaxed and well oriented (lack of mosaicity). No peak corresponding to SiGe is visible in either RSM, which is not surprising due to the sub-10-nm SiGe thickness and expected broadening due to composition and strain variations in the layer. Broadening in the ω (radial) direction is associated with slight variation in GaAs layer tilt. The samples show noticeable omega broadening but only in the (440) GaAs/Ge film peak, not the Si substrate. This indicates the broadening is the result of variation in layer itself and not wafer bowing. Figure S1e displays line scans in the ω direction across the GaAs/Ge(440) peaks for the two samples, corresponding to a Lorentz FWHM of 0.26 and 0.18 for samples A and B, respectively. The sharper peak observed for sample B is consistent with the TEM results, showing increased uniformity and lower defects in the GaAs layer for sample B.

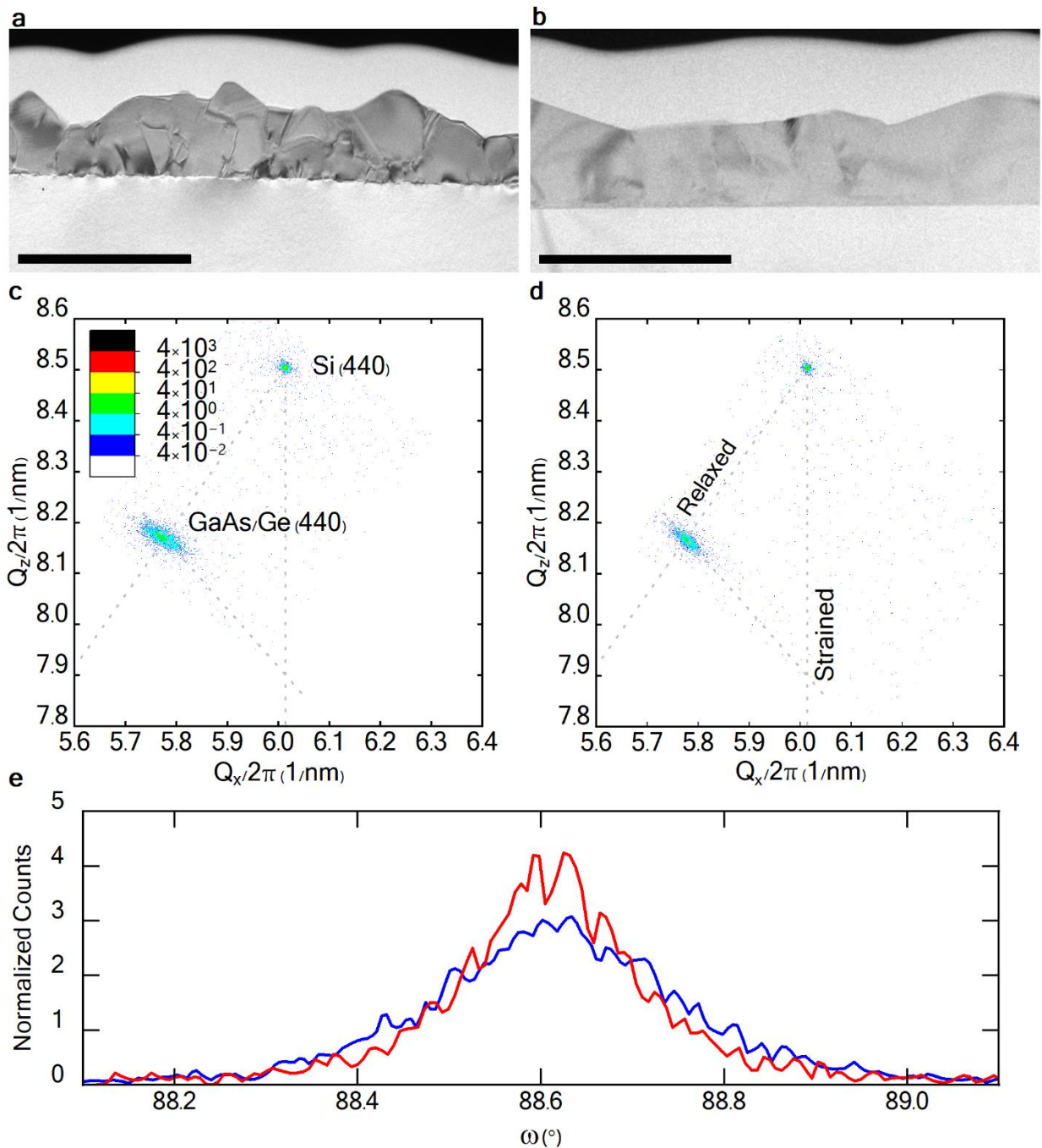

Figure S1: a-b) a-b) Cross-sectional TEM micrographs of samples A and B, respectively, with scale bars of 1 μm. c-d HR-XRD RSMs around the (440) peaks of samples A and B, respectively. The 440 peaks for both Si and GaAs/Ge are visible in the RSM, as indicated in (c). Qz is oriented along the [111] direction and Qy along [112]. The logarithmic intensity scale in figure (c) corresponds to both maps. Gray dotted lines in c and d indicate the lines along which fully strained and fully relaxed layers would lie, along with a line connecting the positions of fully strained and relaxed GaAs/Ge, as annotated in (d). (e) GaAs/Ge peaks from (c) and (d)—samples A and B—plotted along ω in blue and red, respectively.

## TEM

Figure S2a presents the atomic-resolution TEM used in Figure 2 of the main text for GPA analysis. Figure S2b shows the FFT of Figure 2a in the main text with the $g_1 = [2\bar{2}0]$ and $g_2 = [111]$ peaks used to generate the GPA. Figures S2c-d are the uncropped GPA images from Figure 2a using g-vectors $g_1$ and $g_2$. The resulting extra contrast of the GPA image corresponds to edge artifacts – this was verified through careful inspection of the atomic-resolution image previously shown, revealing defects only at the locations shown in the cropped version of the GPA.

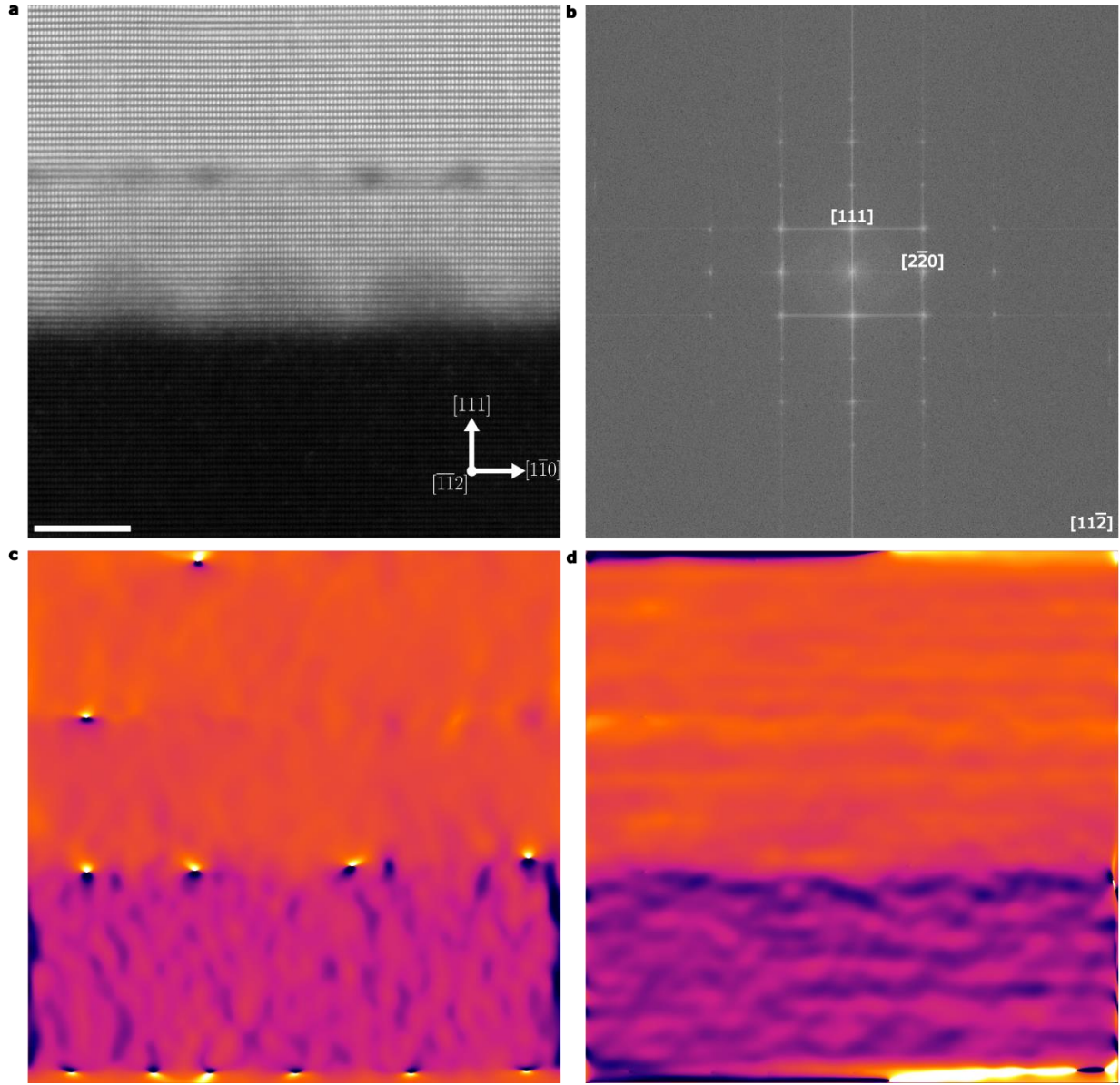

*Figure S2: a) Atomic-resolution TEM from Figure 2 used for GPA analysis. b) Fast Fourier transform (FFT) of Figure 2a from the main text/subfigure a with annotations. c-d) Uncropped GPA images each showing edge artifacts, with $\epsilon_{xx}$ and $\epsilon_{yy}$ GPA strain maps corresponding to c-d, respectively.*